\documentclass[%
aps, 
twocolumn,
superscriptaddress,
bibnotes,
 amsmath,amssymb,
 aps,
prb,
numerical
]{revtex4-2}

\usepackage{comment}
\usepackage{natbib}
\usepackage{todonotes}

\usepackage{etoolbox}
\usepackage[utf8]{inputenc}
\usepackage{siunitx}
\usepackage{array}
\usepackage{physics}
\usepackage{dsfont}
\usepackage[english]{babel}			% English
\usepackage{xcolor}
\usepackage{graphicx}				% images
\usepackage{amssymb,amsmath}		% math
\usepackage{url}					% clickable links
\usepackage{marvosym}				% symbols
\usepackage[T1]{fontenc}            %
\usepackage{wrapfig}				% wrapping text around figures
\usepackage{charter} 				% Charter font for main content
\usepackage{blindtext}				% dummy text
\usepackage{datetime}				% custom date
\usepackage{lipsum}                 % Random latina passages
\usepackage{float}                  %float 
\usepackage{tabularx}
\usepackage{siunitx}
\usepackage{dcolumn}% Align table columns on decimal point
\usepackage{bm}% bold math
\usepackage{todonotes}
\usepackage{hyperref}
\hypersetup{colorlinks=true,allcolors=blue}

\newcommand{\UIBK}{Institut f{\"u}r Experimentalphysik, Universit{\"a}t Innsbruck, 6020 Innsbruck, Austria}
\newcommand{\JKU}{Institute of Semiconductor and Solid State Physics, Johannes Kepler University Linz, 4040 Linz, Austria}
\newcommand{\Camm}{Cavendish Laboratory, JJ Thomson Avenue, University of Cambridge, CB3 0HE Cambridge, UK}
\newcommand{\IITD}{Department of Electrical Engineering, Indian Institute of Technology Delhi, Delhi 110016, India}
\newcommand{\Ucamp}{Universidade Estadual de Campinas, Campinas- SP, 13083-970, Brazil}

\begin{document}

\title{Passive Demultiplexed Two-photon State Generation from a Quantum Dot}

\author{Yusuf Karli}
\thanks{These authors contributed equally}
\affiliation{\UIBK}
\affiliation{\Camm}
\author{Iker Avila Arenas}
\thanks{These authors contributed equally}
\affiliation{\UIBK}
\author{Christian Schimpf}
\affiliation{\Camm}
\author{Ailton Jose Garcia Junior}
\affiliation{\JKU}
\author{Santanu Manna}
\affiliation{\JKU}
\affiliation{\IITD}
\author{Florian Kappe}
\affiliation{\UIBK}
\author{Ren\'e Schwarz}
\affiliation{\UIBK}
\author{Gabriel Undeutsch}
\affiliation{\JKU}
\author{Maximilian Aigner}
\affiliation{\JKU}
\author{Melina Peter}
\affiliation{\JKU}
\author{Saimon F Covre da Silva}
\affiliation{\JKU}
\affiliation{\Ucamp}
\author{Armando Rastelli}
\affiliation{\JKU}
\author{Gregor Weihs}
\affiliation{\UIBK}
\author{Vikas Remesh}
\affiliation{\UIBK}

\makeatletter
% \frontmatter@RRAP@format is responsible for the parentheses
\patchcmd{\frontmatter@RRAP@format}{(}{}{}{}
\patchcmd{\frontmatter@RRAP@format}{)}{}{}{}
\renewcommand\Dated@name{}
\makeatother

\date{Date: \today \\ \phantom{XXX} E-mail: yk441@cam.ac.uk - vikas.remesh@uibk.ac.at}

\begin{abstract}
%High-purity multi-photon states are key elements to realize photonic quantum computing. Among the currently developed infrastructure, semiconductor quantum dots hold great promise toward realizing a scalable and deterministic source of multi-photon states. Yet, identifying a suitable optical excitation method is crucial to extracting the full potential of quantum dot sources. On the other hand, existing methods of multi-photon state generation rely on active polarization-modulating elements. Here we present a technique to generate 2-photon states based on modified Two-Photon Excitation (TPE) of a quantum dot. Following the biexciton state generation via TPE, a second, stimulation laser pulse triggers the biexciton to the exciton state, with the polarization of the emission determined by the stimulation pulse. Using a sequence of horizontal (H) and vertical (V) stimulation laser pulses we then demonstrate a controlled H-V-H-V emission sequence of high-purity single photons. Our results offer a simplified and cost-effective approach at high rates for multi-photon state generation directly from the quantum dot.
High-purity multi-photon states are essential for photonic quantum computing. Among existing platforms, semiconductor quantum dots offer a promising route to scalable and deterministic multi-photon state generation. However, to fully realize their potential we require a suitable optical excitation method. Current approaches of multi-photon generation rely on active polarization-switching elements (e.g., electro-optic modulators, EOMs) to spatio-temporally demultiplex single photons. Yet, the achievable multi-photon rate is fundamentally limited by the switching speed of the EOM. Here, we introduce a fully passive demultiplexing technique that leverages a stimulated two-photon excitation process to achieve switching rates that are only limited by the quantum dot lifetime. We demonstrate this method by generating two-photon states from a single quantum dot without requiring any active switching elements. Our approach significantly reduces the cost of demultiplexing while shifting it to the excitation stage, enabling loss-free demultiplexing and effectively doubling the achievable multi-photon generation rate when combined with existing active demultiplexing techniques.
%These results mark a crucial step toward scalable quantum light sources for photonic quantum computing.
\end{abstract}

\maketitle

\section{Introduction}
Photonic quantum computing offers a unique advantage over other quantum platforms due to the long coherence time of photons, enabling robust quantum communication, quantum information processing, and quantum simulations \cite{aspuru2012photonic,sparrow2018simulating}. A critical requirement for these applications is the reliable generation of high-purity multi-photon states, i.e., $n$ indistinguishable photons in $n$ spatial modes -- which serve as fundamental building blocks for quantum algorithms, error correction, quantum simulations, and advanced photonic networks \cite{wang2017high, zhong2020quantum, kimble2008quantum}. Multi-photon states are also essential for probing quantum optical phenomena such as multi-photon interference \cite{menssen2017distinguishability,faleo2024entanglement}. The most widely used sources to produce multi-photon quantum states are the ones relying on parametric down-conversion or four wave mixing in nonlinear crystals. However, the scalability here is limited, due to the probabilistic nature of photon emission and the required resource overhead for computing and boson sampling applications. 

Semiconductor Quantum Dots (QDs), recognized as deterministic and bright sources of exceptionally pure single-photon Fock states \cite{schweickert2018demand, heindel2023quantum, frick2023single, karli2024controlling} hold great promise in these applications. Several optical schemes have been developed to trigger QD sources, achieving near-unity state-preparation efficiency \cite{bracht2021swing,karli2022super,wilbur2022notch}, controlled polarization \cite{Thomas2021b,Somaschi2016}, improved photon quality \cite{sbresny2022stimulated}, and robustness for practical applications\cite{RemeshCFBG,karli2024robust,ramachandran_experimental_2021,kappe2024chirped}. The QD-based sources are well-compatible with various photonic integrated platforms, and are highly tunable by controlling the growth processes \cite{kuroda2013symmetric,juska2013towards,versteegh2014observation}, enabling on-chip sources for advanced photonic quantum computing devices. Furthermore, via strain-tuning \cite{seidl2006effect,zhang2015high,trotta2015energy}, external electric or magnetic fields \cite{stevenson2006semiconductor,muller2009creating,kowalik2005influence}, integrating into photonic cavities one can control the spectro-temporal indistinguishability between photons emitted from multiple QDs. Recently, several research works have demonstrated interference between photons emitted from remote QDs \cite{zhai2022quantum,reindl_phonon-assisted_2017}, albeit with heavy experimental overhead, i.e., $n$ cryostats for $n$-photon interference. 

The alternate approach found in literature is to use the single-photon stream from a \textit{single} QD and spatio-temporally demultiplex them, typically by employing electro-optic modulators (EOMs) \cite{munzberg2022fast} or reconfigurable on-chip waveguides \cite{lenzini2017active}. This is the so-called \textit{active} demultiplexing scheme. Based on this technique, several groups have successfully demonstrated multi-photon quantum interference experiments \cite{cao2024photonic, chen2024heralded, wang2019boson, sund2023high, maring2024versatile}. 
Typically, EOM-based demultiplexing works as follows: a resonant EOM is operated at half the photon generation rate (i.e., the laser repetition rate) to ensure that two consecutively emitted photons are routed into two orthogonal polarization states \cite{munzberg2022fast}.  To scale it to \textit{n} spatial modes, one needs a total of \textit{n-1} EOMs. The critical limitation in generating an $n$-mode photon state via spatio-temporal demultiplexing system is the first and the fastest EOM-- often the most inaccessible one. Though high-speed fiber-based EOMs are commercially available, they incur few-dB losses.  

\begin{figure*}
    \centering
    \includegraphics[width=\textwidth]{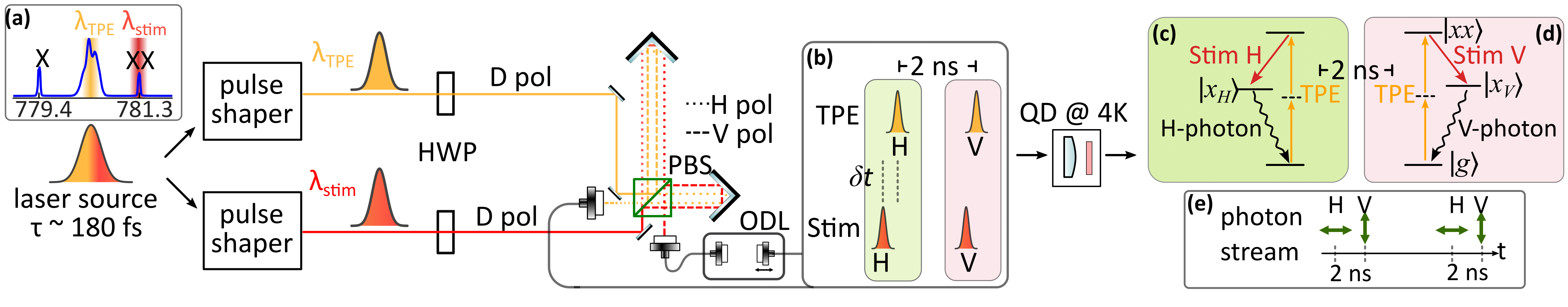}
    \caption{\textbf{Conceptual sketch of our passive demultiplexing}: Two 4$f$ pulse shapers enable the preparation of the TPE (orange-shaded) and sTPE (red-shaded) pulses from a broadband femtosecond laser source. The inset (a) denotes the emission spectrum under TPE, where the labels $\lambda_{\text{TPE}}$ and $\lambda_{\text{stim}}$ denote the spectral ranges for TPE and sTPE respectively. The half-wave plates (HWP) ensure diagonal polarization for both pulses before they enter a pulse-pair generator. The latter is equipped with a polarizing beamsplitter (PBS) to produce a time-delayed sequence of H-V polarized TPE and stim pulses (inset (b)). To precisely set the time delay of the stim pulse from the TPE pulse, we use a fiber-optic delay line (ODL). These pulses then excite the QD kept in a cryogenic microscope at \SI{4}{K}. The insets (c) and (d) present the QD level diagrams corresponding to the H and V-polarized sTPE schemes. The inset (e) depict the obtained H-V...H-V photon stream.}
    \label{fig:setup}
\end{figure*}
In this work we introduce an active-element-free approach to directly trigger a bi-modal photon state from a QD source. Our method relies on Two-Photon Excitation (TPE) of the QD followed by precisely timed and polarization-tailored stimulation (stim) pulse-pair to achieve high-purity, bi-modal photon emission. 
We have chosen TPE to excite the QD because, firstly, it is well known for producing record-high purity single photons via the biexciton ($\ket{xx}$)-exciton ($\ket{x}$)-ground state ($\ket{g}$) cascade \cite{hanschke_quantum_2018,schweickert2018demand}.
Most importantly, it intrinsically enables polarization-correlated photon emission, i.e., in horizontal (H) and vertical (V) polarization states \cite{benson_regulated_2000, akopian_entangled_2006}, due to the angular momentum properties of the biexciton states.
Finally, because the excitation laser energy is different from the emitted photon energy, one does not require a challenging cross-polarization filtering, enabling us to double the photon rate in a given polarization. Following the near-unity preparation of the biexciton state, we send the stim pulse, which is energetically tuned to the biexciton-exciton transition. The polarization of the stim pulse dictates the polarization of the subsequent exciton-ground state photon emission \cite{akimov2006stimulated}. We refer to this technique as stimulated TPE (sTPE), which has recently been employed to produce high-indistinguishability photons \cite{sbresny2022stimulated, wei2022tailoring} with a tailored degree of photon number coherence \cite{karli2024controlling}. In Figure \ref{fig:setup} inset we summarize the scheme, where the green and the magenta coloured boxes respectively denote the H and V photon generation processes. It is important to note that the arrival time of the stim pulse with respect to the TPE pulse is approximately \SI{6}{ps} and is determined experimentally for optimal performance. 
In other words, exciting a single QD using sTPE pulse pairs (H and V) gives rise to a deterministic stream of high-purity single-photon pairs (H and V). Most importantly, our simplified and cost-effective approach eliminates the need for the fastest EOM in a spatio-temporal demultiplexing scheme, and therefore is not switching-rate limited. This means that by tailoring the repetition rate of the laser source, one can trigger multi-photon states directly from QDs, limited \textit{only} by the exciton state lifetime.

\section{Passive demultiplexing scheme}
The schematic of our passive multi-photon generation is described in Figure \ref{fig:setup}. It consists of laser pulse shaping, polarization-tailored pulse-pair generation, and cryogenic microscope hosting the QD source. In the first stage, a Ti:Sapphire laser source (Chameleon Ultra II, Coherent) is tuned to \SI{782}{nm} and coupled to two independent 4f pulse shapers \cite{kappe2024chirped, karli2024robust} allowing spectral shaping of the TPE (at \SI{780.4}{nm}) and stim pulses (at \SI{781.3}{nm}), whose polarizations are set to diagonal using half-wave plates. We use variable optical attenuators (VOA, V800PA, Thorlabs) to independently control the intensities of the TPE and stim pulses. Both pulses enter a pulse-pair generator, where the path length differences between the two arms are adjusted to produce a pulse-pair of \SI{2}{ns} time delay. The choice of the \SI{2}{ns} time delay $\delta t$ is based on the reported lifetime of the QD \cite{karli2024controlling,karli2024robust}, such that $ \SI{2}{ns}  >> \tau$ where $\tau$ is the lifetime of the QD, thereby avoiding the chances of reexcitation before the state decays. Note that we use a polarizing beam splitter (PBS) in the pulse-pair generator to produce pairs of H and V polarized TPE and stimulation pulses. The polarization of the TPE pulse is not relevant in preparing the $\ket{xx}$, as any linear polarization state can do so. To control the time difference between the identically polarized duo of TPE and stim pulse we use a fiber optic delay line (ODL-300, OZ Optics). Afterwards, the TPE and stim pulses are directed to the cryostat equipped with an asphere lens (Edmund Optics NA 0.77), hosting the QD sample. Our sample consists of a GaAs/AlGaAs QDs grown using molecular beam epitaxy with local droplet etching \cite{da2021gaas}, emitting around \SI{780}{nm} (see details in SI). The QDs are embedded into a p-i-n diode structure to control the local charge environment. The collected single-photon emission at exciton energy (X) from the QD is spectrally filtered using a home-built monochromator consisting of narrow-band notch filters (FWHM \SI{0.3}{nm}, Optigrate) and separated into H and V components, and detected using superconducting nanowire single-photon detectors (Eos, Single Quantum). To determine the photon quality, we employ a Hanbury Brown and Twiss (HBT) setup measuring $g^{(2)}_{(0)}$ and a Hong-Ou-Mandel (HOM) setup measuring the indistinguishability of successively emitted photons. 
%Further details on the setup are described elsewhere \cite{karli2024controlling}.  

\section{Experimental Results}
\begin{figure} 
\centering 
\includegraphics[width=\linewidth]{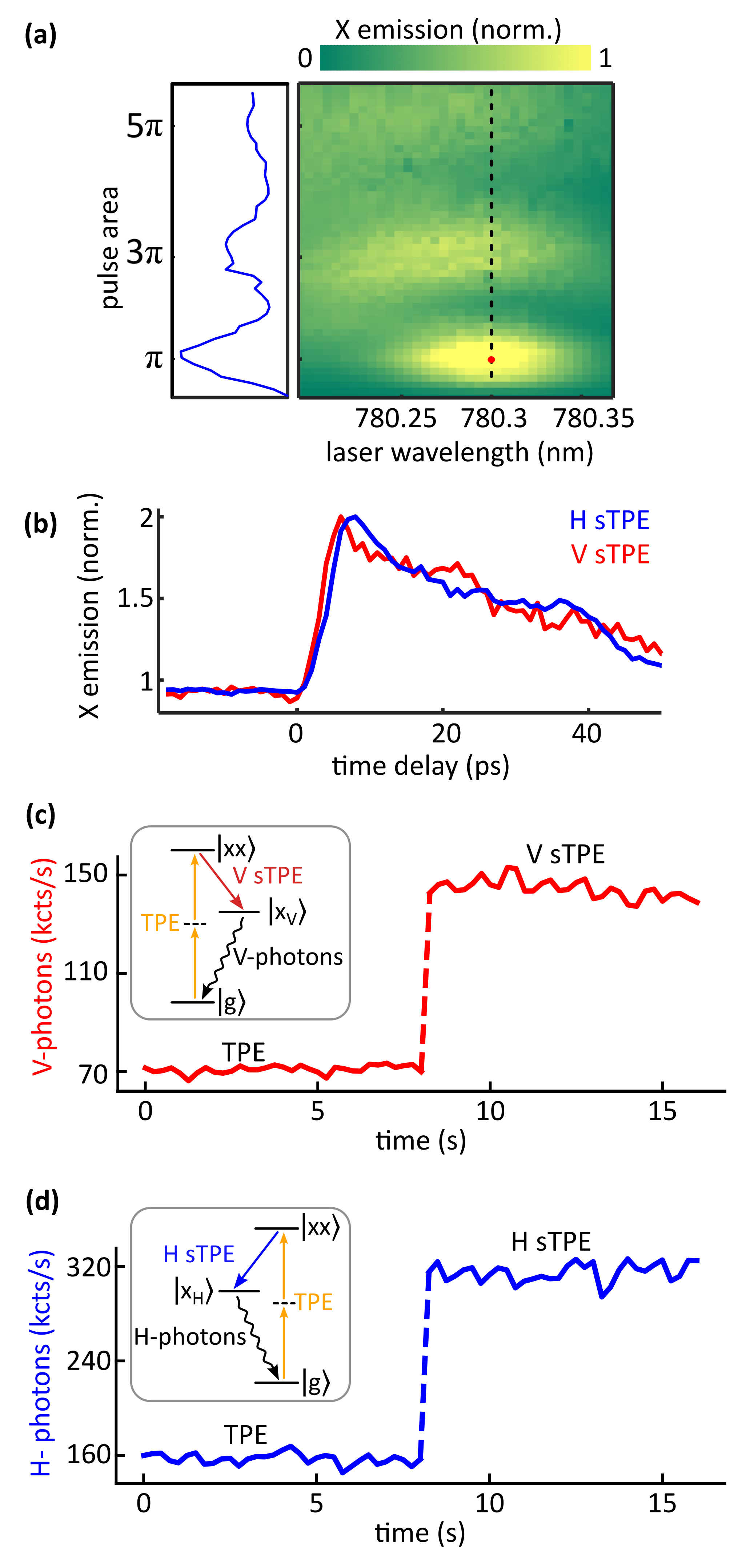} 
\caption{\textbf{Coherent control and deterministic stimulation of polarized photon states}: (a) Results of the Rabi rotation experiment as a function of TPE wavelength (x-axis) and pulse area (y-axis). The dashed line indicates the TPE resonance and red dot indicates the $\pi$ pulse condition where the QD source is operated. The left inset depicts the line cut along the dashed line. (b) Deterministic stimulation of polarized photon states via sTPE: recorded X emission as a function of the time delay between the TPE and stim pulses for H (blue curve) and V (red curve) polarizations. (c), (d) Representative time traces of photon counts recorded under TPE (seconds 0-8) and sTPE (seconds 8-16) for both polarizations. Insets denote the respective stimulation and emission processes.} 
\label{fig:fig:stim_proof} 
\end{figure}

To enable TPE to $\ket{xx}$ we tune the laser pulse to the TPE resonance at \SI{780.3}{nm}. This results in biexciton to exciton emission at \SI{781.3}{nm} (hereafter called XX photon) and exciton to ground state emission at \SI{779.4}{nm} (called X photon). To verify the coherent control of $\ket{x}$ we perform a resonant Rabi rotation experiment by sweeping the TPE laser power and wavelength, while recording the X photon counts. The results are presented in Figure \ref{fig:fig:stim_proof} (a), which shows Rabi rotations in X population. 

Next, by setting both outputs of the pulse pair generator to be the TPE and sTPE $\pi$ powers, we optimize the efficiency of sTPE. To this end, the orthogonally polarized pairs (time delay of \SI{2}{ns}) of TPE pulses and stim pulses are sent to the QD, such that the stim pulse is slightly delayed by $\delta t$ from its TPE counterpart. It is known that a precise setting of $\delta t$ is important to ensure a near-unity stimulation of the desired polarization channel, i.e., effectively doubling the photon rate in this polarization and obtain the highest indistinguishability of emitted X photons \cite{karli2024controlling,sbresny2022stimulated,wei2022tailoring}. To verify the former, and to maximize the photon count rate under sTPE, we turn on the V-polarized stim pulse and perform a time delay scan. The results are presented in Figure \ref{fig:fig:stim_proof} (b). At $\approx$ \SI{6}{ps} time delay, we observe that the photon counts reach twice that under TPE, clearly establishing the successful stimulation to V-polarization (red curve). Following this, we repeat the identical procedure for the H-polarized scenario and obtain nearly double the photon emission (blue curve). 

Subsequently, we establish the sequential generation of H and V-polarized photon states, we proceed as follows: the first TPE pulse excites the $\ket{xx}$, followed by the V-polarized stim pulse deterministically triggering the V-polarized X emission. Afterwards, the second TPE pulse prepares the $\ket{xx}$, followed by the H-polarized stim pulse generating the corresponding X emission. This is represented in Figure \ref{fig:fig:stim_proof} (c) and (d), as two exemplary \SI{8}{s} time traces of photon emission at TPE and sTPE conditions for both polarizations. This sequence is repeated at every \SI{12.5}{ns} interval, corresponding to the original repetition rate of the laser source, resulting in an H-V H-V ... sequence of photon states.

\begin{figure*} 
\centering 
\includegraphics[width=\textwidth]{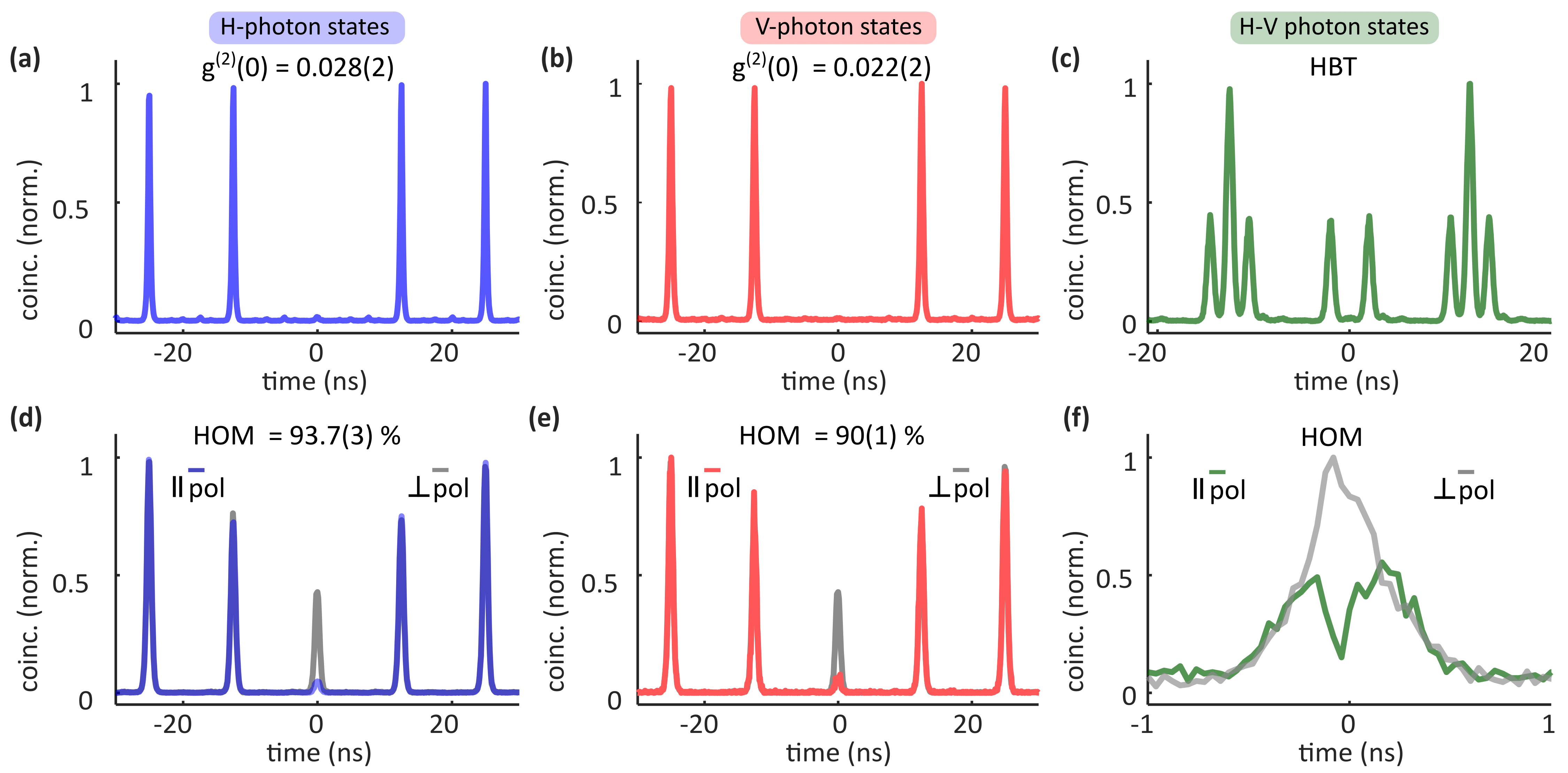} 
\caption{\textbf{Quality of the generated 2-photon states}: Panels \textbf{(a)} and \textbf{(b)} represent the measured $g^{(2)}_{(0)}(H)$ (blue) and $g^{(2)}_{(0)}(V)$ (red), of the H-polarized and V-polarized photon states respectively. Panels \textbf{(d)} and \textbf{(e)} represent the results of HOM coincidence measurements. Here, blue (gray) and red (gray) curves represent the coincidences recorded for parallel and orthogonal polarizations in respective cases. Panel \textbf{(c)} (green curve) shows the recorded coincidences in the HBT experiment while collecting both H and V photons. Panel \textbf{(f)} (green curve) shows the recorded coincidences in the HOM experiment for combined H-V photon states, where the gray curve denotes the coincidences recorded for orthogonal polarization configuration.} \label{fig:HBT_HOM}
\end{figure*}
To finally characterize the nature of the emission, we also record the X emission lifetimes under TPE and then under sTPE cases (see SI for results). The estimated lifetime for the exciton photons is found to be \SI{175(4)}{ps}, consistent with previous reports on the GaAs QDs.

To characterize the quality of the obtained two independent, orthogonally polarized single photons, we first measure the second-order correlation function $g^{(2)}(0)$ using HBT setup, with the output ports coupled to SNSPDs. The results are presented in Figures \ref{fig:HBT_HOM}, panels (a) and (b). The $g^{(2)}(0)$ values are determined as the ratio of the mean area of the first side peaks at \SI{12.5}{ns} to the central peak area, using an integration window of \SI{1}{ns} for each peak. The calculated values are $g^{(2)}(0) = 0.028(2)$ for H-polarized photons and $g^{(2)}(0) = 0.022(2)$ for V-polarized photons, which confirm the excellent single-photon properties of the generated states. The uncertainties are obtained assuming Poissonian statistics for the photon counts.

Next, we measure the single-photon indistinguishability, using the HOM setup. The polarization in both arms of the setup is controlled individually, and their time delay difference is set as $\pm\SI{12.5}{ns}$. This enables a comparison between the cross-polarized (distinguishable) case and the co-polarized (indistinguishable) case. In Figures \ref{fig:HBT_HOM} (d),(e) we present the recorded coincidence histograms, where the blue and red curves denote the HOM data for H and V polarized photons in the co-polarized cases and the gray curves represent the cross-polarized case. The raw HOM indistinguishability values were calculated by comparing the area under the central peak of each co-polarized case to the cross-polarized case, for \SI{1}{ns} time window and Poissonian statistics for the uncertainties. 
HOM visibilities are corrected using the formula \cite{munzberg2022fast}:

\begin{equation}
\text{HOM} = \frac{V_{\text{raw}} + g^{(2)}(0)}{1 - g^{(2)}(0)} \cdot \frac{R^2 + T^2}{2 R T}
\end{equation}
where \( R (0.47) \) and \( T (0.53) \) represent the measured values of the reflection and transmission coefficients of the beam splitter, respectively.
We measured raw HOM visibilities ($V_{\text{raw}}$)  of $87.6(3)\%$) for H-polarized photons and 
$84.0(1)\%$ for V-polarized photons. After correcting, the HOM visibilities increase to $93.7(3)\%$ for H-polarized and 
$90(1)\%$ for V-polarized photons, respectively.
These high HOM visibilities confirm the high degree of indistinguishability of the generated single photons in both polarizations. 

Following the photon quality measurements, we now extend our method to generate multi-photon states. Initially, the HBT experiment is repeated while simultaneously recording the single-photon emission of both H and V cascades. In Figure \ref{fig:HBT_HOM} (c) we present the resulting histogram which clearly shows multiple peaks: at short time delays, we observe two peaks while for larger delays we observe triplets, as expected for two independent single photon sources. We then look for HOM visibility between H and V-polarized photon states in a modified HOM setup where the time difference between the co- and cross-polarized arms is set to \SI{2}{ns}. This corresponds to the time delay of photon emission, as decided by the pulse-pair generator. The recorded coincidences, obtained using SNSPDs and a time tagger, are shown in Figure \ref{fig:HBT_HOM} (f), where the green and gray curves represent the co- and cross-polarized cases respectively. The computed raw indistinguishability is 28(2)\%.
\begin{figure}
\begin{center}
\includegraphics[width=\linewidth]{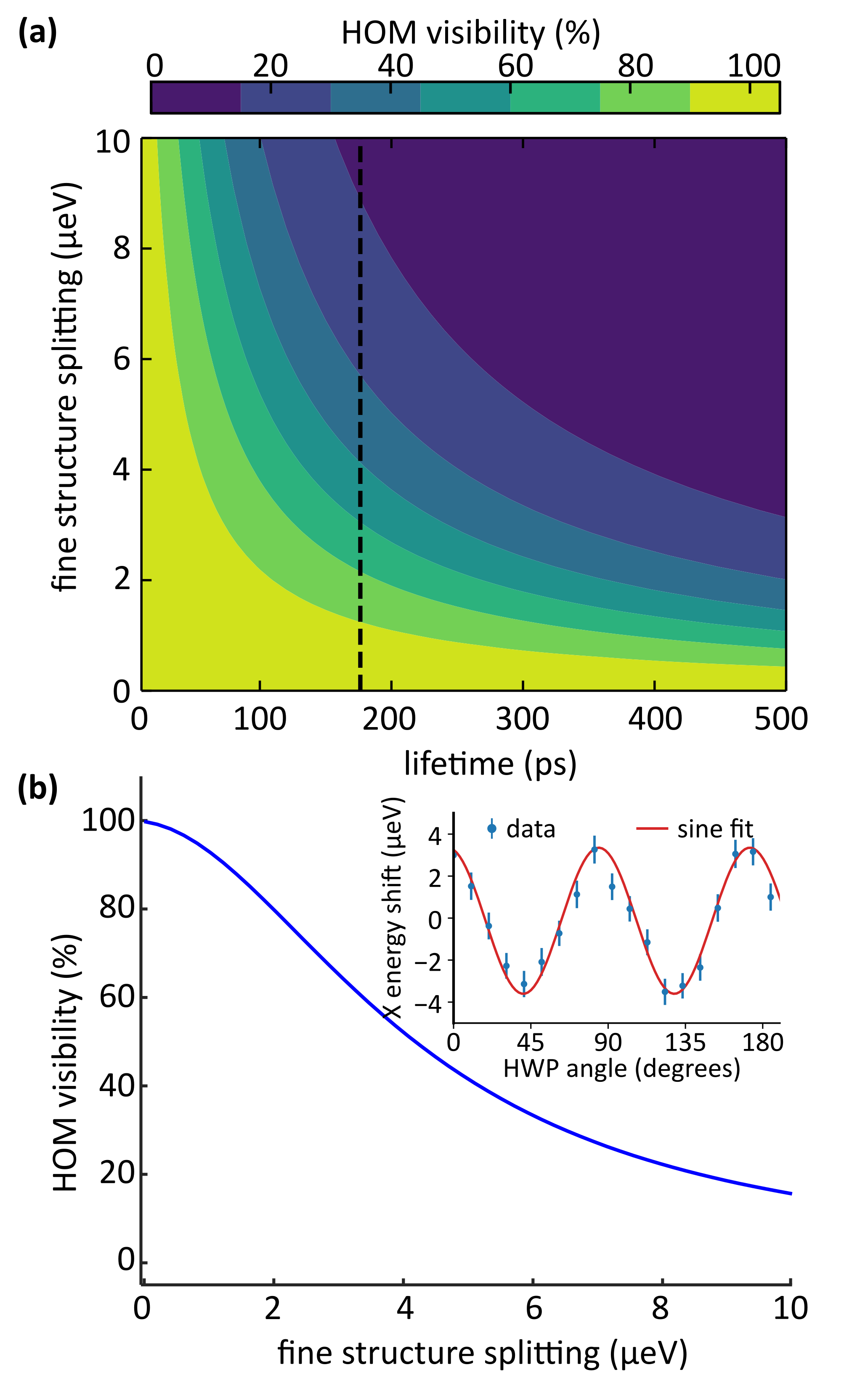}
\end{center}
\caption{\textbf{Indistinguishability limit}: \textbf{(a)}: The simulated HOM visibility as a function of lifetime and fine structure splitting of the QD. The dashed line represents the scenario for the QD measured. \textbf{(b)}: The line-cut of \textbf{a}, corresponding to the lifetime of \SI{170}{ps}. Inset shows the result of FSS measurement on the considered QD.}
\label{fig:V}
\end{figure}

\section{Discussion}
To understand the limit of achievable indistinguishability in the generated 2-photon states, we start by looking into the influence of the fine structure splitting (FSS) of the QD \cite{bayer2002fine}. For the investigated QD, we measured an FSS of \SI{7.0(3)}{\micro\electronvolt}, as presented in Figure \ref{fig:V} (inset). To estimate the effect of FSS on the combined indistinguishability between H- and V-polarized photon emissions, we use the model introduced in Ref. \cite{kambs2018limitations}. 
For a difference in central frequency between H- and V-polarized photon emission of $\delta\nu = FSS/h$ and for a radiative lifetime of $T_1$, the HOM visibility V is given by 

\begin{equation}
    V = \frac{\text{Re}[w(z)]}{\sqrt{2\pi} \Sigma\, 2T_1},\quad  z = \frac{2\pi \delta\nu + i\gamma}{2\pi\sqrt{2}\Sigma},
\end{equation}
with $w(z)$ being the Faddeeva function, $\Sigma$ the standard deviation of the inhomogeneous broadening from spectral wandering, and $\gamma$ the dephasing. If pure dephasing like phonon broadening is neglected, $\gamma = 1/T_1$. Figure \ref{fig:V} shows the behavior of $V$ for $\Sigma\rightarrow 0$ as a function of $T_1$ and $\delta \nu$, and Figure \ref{fig:V} (b) shows the line-cut corresponding to $T_1=\SI{170}{ps}$. As our QDs are embedded in a p-i-n diode structure, spectral wandering is nearly absent. For the FSS of \SI{7}{\micro\electronvolt}, the maximum expected indistinguishability between the H- and V-polarized photons is found to be 25\%, which is in agreement with the experimental value. 

The indistinguishability between the H and V-polarized photon streams we generated is fundamentally limited by the FSS and lifetime of the QD. 
This opens up two distinct use cases for our technique. Firstly, where the QD does not provide high indistinguishability, it enables the generation of two distinct streams of single photons (H and V) from a single laser and a single QD, overcoming the need for multiple sources or multiple QDs and cryostats holding the QDs. One potential application is simultaneous secure quantum communication with two different parties without sacrificing the bit rate. 
Importantly, the more critical use case arises when high indistinguishability of H-V photons is available. In this case, our approach significantly enhances the existing multiplexing schemes by doubling the maximum achievable rate \cite{chen2024heralded,cao2024photonic}. This capability could dramatically improve the performance of quantum algorithms, simulations, and error correction protocols, providing a key advantage for scalable quantum technologies.
We identify that higher indistinguishability could be achieved in two ways: lower FSS or lower lifetimes. While the former is achievable via optimized growth processes \cite{kuroda2013symmetric,juska2013towards,versteegh2014observation}, strain-tuning \cite{seidl2006effect,zhang2015high,trotta2015energy}, applying external magnetic fields \cite{stevenson2006semiconductor} or electric fields \cite{muller2009creating,kowalik2005influence}, the latter can be achieved by integrating QDs in engineered photonic cavities such as circular Bragg gratings \cite{Rickert_2024_pigtail}. 

Several further interesting recent advancements also align with our approach. For instance, compact, few-element GHz laser sources are now feasible \cite{ostapenko2022three}, and on-chip Ti:Sapphire lasers are approaching commercialization \cite{yang2024titanium, Schlehahn2015, Mangold2014}. Banking on this, our technique would help generate multi-photon states at GHz clock rate from a QD. In tandem, recent demonstrations of fiber-coupled single-photon sources \cite{Rickert_2024_pigtail, Northeast2021_fibernanowire}, and techniques for robust, plug-and-play optical excitation of QD sources based on chirped fiber Bragg gratings \cite{RemeshCFBG,karli2024robust}, suggest that compact, high-repetition-rate, and large-scale field-deployable QD photon sources are forthcoming, which would benefit from an elegant all-optical technique described in this work.

\section{Conclusion}
%To summarize, we have demonstrated a fully passive demultiplexing method to generate multi-photon states from a QD. Our technique relies on two-photon excitation of a QD, followed by a precisely timed, polarization-tailored stimulation pulse. Using this approach, we successfully generated a stream of H and V-polarized photons directly from the QD. This method is conceptually similar to exciting a QD source at \SI{160}{MHz}, for a targeted spatio-temporal demultiplexing using active elements at a rate of \SI{80}{MHz}--something that exceeds the current state-of-the-art. Despite the anticipated development of faster EOMs based on advancements in ultrafast switching using nonlinear materials, it is crucial to explore flexible, scalable, cost-effective alternatives such as the one demonstrated in our work.

To summarize, we have demonstrated a fully passive demultiplexing method to generate two-photon states from a QD. Our technique relies on two-photon excitation of a QD, followed by a precisely timed, polarization-tailored stimulation pulse. Using this approach, we successfully generated a stream of H- and V-polarized photons directly from the QD. This method is conceptually similar to exciting a QD source at \SI{160}{MHz} and then performing targeted spatio-temporal demultiplexing with active elements at \SI{80}{MHz}, which pushes the limits of current state-of-the-art  free-space EOM technology. In contrast, switching rate of our passive multiplexing technique is limited only by the exciton state lifetime. Since the QD lifetimes allow for a gigahertz-rate excitation cycle, this approach removes not only the need for a fast EOM but also reduces photon loss by moving the demultiplexing step to the excitation process—thereby overcoming one of the crucial obstacles on the path to photonic quantum computing.

\section{Acknowledgements}
YK, IAA, RS, FK, GW, GU, MA,  SM, AR and VR acknowledge the financial support through the Austrian Science Fund (FWF) projects with Grant DOIs 10.55776/TAI556 (DarkEneT), 10.55776/W1259 (DK-ALM Atoms, Light, and Molecules), 10.55776/FG5, 10.55776/I4380 (AEQuDot), 10.55776/COE1 (quantA), and the infrastructure funding from FFG (HuSQI, grant number FO999896024). AR acknowledges the Linz Institute of Technology (LIT) and the European Union's Horizon 2020 research, and innovation program under Grant Agreement Nos. 899814 (Qurope), 871130 (ASCENT+), and the QuantERA II Programme (project QD-E-QKD). CS acknowledges the Austrian Science Fund project 10.55776/J4784. For open access purposes, the authors have applied a CC BY public copyright license to any author-accepted manuscript version arising from this submission.
 
\section*{References}
\bibliographystyle{apsrev4-2}
\bibliography{Main}

\end{document}